\documentclass[aps,twocolumn,superscriptaddress,prd,floatfix]{revtex4}
\usepackage{graphicx,subfigure,amsmath}
\newcommand{\ben}{\begin{equation}}
\newcommand{\een}{\end{equation}}
\newcommand{\bea}{\begin{eqnarray}}
\newcommand{\eea}{\end{eqnarray}}
\newcommand{\ba}{\begin{array}}
\newcommand{\ea}{\end{array}}
\newcommand{\bit}{\begin{itemize}}
\newcommand{\eit}{\end{itemize}}

\def\quarter{\frac{1}{4}}

\begin{document}

\title{Semilocal Cosmic String Networks}

\author{Ana Ach\'ucarro}
\affiliation{Instituut-Lorentz, Universiteit Leiden,
P.\,O. Box 9506, 2300 RA Leiden, The Netherlands}
\affiliation{Department of Theoretical Physics, University of the Basque
Country UPV-EHU, 48080 Bilbao, Spain}

\author{Petja Salmi}
\affiliation{Instituut-Lorentz, Universiteit Leiden,
P.\,O. Box 9506, 2300 RA Leiden, The Netherlands}
\affiliation{Department of Physics \& Astronomy, 
University of Sussex, Falmer, Brighton BN1 9QH, UK}

\author{Jon Urrestilla}
\affiliation{Department of Physics \& Astronomy, 
University of Sussex, Falmer, Brighton BN1 9QH, UK}
\affiliation{Institute of Cosmology, Department of Physics and 
Astronomy, Tufts University, Medford, MA 02155, USA}

\date{\today}
\begin{abstract}

  We report on a large scale numerical study of networks of semilocal
  cosmic strings in flat space in the parameter regime in
  which they are perturbatively stable. We find a population of
  segments with an exponential length distribution and indications of
  a scaling network without significant loop formation.  Very deep in
  the stability regime strings of superhorizon size grow rapidly and
  ``percolate'' through the box. We believe these should lead at late
  times to a population of infinite strings similar to topologically
  stable strings.  However, the strings are very light; 
  scalar gradients dominate the energy density and the network has
  thus a global texture-like signature. As a result, the observational
  constraints, at least from the temperature power spectrum of the
  CMB, on models predicting semilocal strings, should be closer to
  those on global textures or monopoles, rather than on topologically
  stable gauged cosmic strings.

\end{abstract}

%\pacs{11.27.+d, 98.80.Cq, 11.15.-q}

\maketitle

\section{Introduction}

The improvement in observational data during the last years,
especially on Cosmic Microwave Background (CMB) radiation, has shown
that topologically stable 
cosmic strings cannot be the dominant
component creating the primordial fluctuations that seed 
large scale structure as we see it in the Universe today. 
Strings could be a subdominant component~\cite{Wyman:2005tu}
but there is as yet no 
evidence for their existence (see~\cite{Sazhin:2003cp} for a
candidate string lensing event that turned out to be two galaxies).
On the other hand, from the theoretical point of view, the formation
of topological defects via the Kibble mechanism~\cite{Kibble:1976sj}
is
inevitable in theories that allow stable defects. The formation of
cosmic strings after inflation appears to be generic both in
supersymmetric grand unified theories breaking down to the Standard
Model~\cite{Jeannerot:1997is,Jeannerot:2003qv} and in brane inflation
models within the framework of fundamental
superstrings~\cite{Sarangi:2002yt}. Therefore, models that do not
predict topologically stable defects at the end of inflation are
somewhat appealing. Recent work~\cite{Urrestilla:2004eh} suggests that
a particular type of non-topological defects, called {\it semilocal}
strings, could form in the early universe both after $D$-term
inflation \cite{Dvali:1994ms} and after brane inflation with
$D3/D7$-branes \cite{Dasgupta:2004dw,Chen:2005ae}.

The semilocal model~\cite{Vachaspati:1991dz} is a minimal extension of
the Abelian Higgs model by a global $SU(2)$ symmetry. With
two equally charged Higgs fields and only one $U(1)$ gauge field,
there are not enough gauge degrees of freedom to cancel all scalar
gradients, even outside the string cores, unless the scalar windings
are correlated.  Semilocal (SL) strings are stable for sufficiently
large gauge coupling~\cite{Vachaspati:1991dz,Hindmarsh:1991jq}, but
due to their non-topological character they appear after the phase
transition as open segments. These are closely related to
electroweak dumbbell configurations 
\cite{Nambu:1977ag,Vachaspati:1992fi,Urrestilla:2001dd}
but
whose ends have long-range interactions that resemble those of global
monopoles~\cite{Hindmarsh:1992yy}, with a force roughly independent of
distance.  Therefore the network of SL strings has several features
that are not present in networks of topologically stable
Abrikosov-Nielsen-Olesen (ANO) strings~\cite{VS} and their
cosmological evolution could be quite
different~\cite{Benson:1993at,Achucarro:1999it}.  Semilocal string
segments can either contract and eventually disappear or grow to join
a nearby segment and form a longer string. Closed loops can be formed
by intercommutations, and also if the two ends of a segment join.  The
obvious question is if at some point this joining of segments can form
infinitely long strings.  If so, the networks present characteristics
intermediate between topological ANO strings and global defects --
both of which show scaling behaviour in an expanding universe
\cite{VS} - and we would like to understand their cosmological impact.

The purpose here is to report on a large scale numerical
study of SL string networks. The non-topological nature of the SL
strings does not permit a Nambu-Goto approach and our results are
based on classical field theory simulations.  Our basic results are
twofold.  We find that SL strings will,
in the right parameter range, form a network that shares features with
a network of ANO strings due to the existence of fast growing,
extremely long percolating strings, which we believe would lead to an
infinite string component at late times (we use the word
``percolating'' in a non-technical sense, to mean that the strings
span the simulation box).  On the other hand, the energy density is
dominated by scalar gradients, like in models for global defects, even
in the case of dense string networks deep in the stable regime.  As
the energy density plays a crucial role for the observable
cosmological consequences, like the effect on the temperature power
spectrum of the CMB, the signature of a SL string network is thus
likely to be very similar to that of global defects.

\section{Semilocal Model}

The action in dimensionless units is given 
by~\cite{Achucarro:1999it}
\begin{eqnarray}
S = 
 \int d^4 x \left[ | D_{\mu} \Phi |^2  - 
\quarter F_{\mu \nu} F^{\mu \nu}  -
\frac{1}{2}{\beta } \Big( \Phi^{\dag} \Phi
- {1} \Big)^2 \right],
\label{semilocalaction}
\end{eqnarray}
where $D_{\mu}\!=\!\partial_\mu - i A_\mu $ and $\Phi^T \! = \! (
\phi_1, \phi_2 )$, is a complex $SU(2)$ doublet.  The only parameter of
the model is the ratio of the scalar and gauge masses, ${\beta} =
m^2_{\rm scalar} / m^2_{\rm gauge}$ (more generally, $\beta \propto
\lambda/ q^2$, where $q$ is the gauge coupling and $\lambda$ the
quartic coupling).

ANO flux tubes embbeded in the SL model 
(e.g. with $\phi_2 \equiv 0$) 
are classical
solutions of~(\ref{semilocalaction}) but they are not
topologically stable because the vacuum manifold (the
three-sphere $|\phi_1|^2 + |\phi_2|^2 = 1$) is simply connected. A
string with $\phi_2 \equiv 0$ can decay into configurations with
$\phi_2 \neq 0$ and  stability is a dynamical question.  It was shown
in~\cite{Vachaspati:1991dz,Hindmarsh:1991jq} that infinite straight
vortices are stable if $\beta<1$ (neutrally stable at $\beta=1$) for
any winding number.  For $\beta>1$ the bare embedded ANO strings 
with unit winding are unstable toward spreading of the
magnetic flux. (A new class of solutions carrying persistent currents
was found \cite{Forgacs:2005sf} whose stability is under
investigation).

The string core carries $U(1)$ magnetic flux. We use this magnetic
field to identify the strings in numerical simulations, as observing
the zeros of scalar fields does not provide an unambiguous detection
method for non-topological defects~\cite{Achucarro:1999it}. 
A SL string can be continuously deformed into ``skyrmion-like''
configurations; an excited string may have no scalar zeroes at the
core but nevertheless it is a linear concentration of energy and 
for cosmological applications it should be 
considered 
a string-like defect. 
Here we
follow the strategy of \cite{Achucarro:1997cx,Achucarro:1998ux}: 
we separately obtain
numerically the profile of the scalar and magnetic field of the
straight ANO string for several values of $\beta$ and use this
information to analyze the data from simulations. 
If the magnetic field strength
at a given point exceeds a certain fraction $f$ 
of the maximum magnetic field of the corresponding ANO
string core, the point is considered to be part of a SL string.
The string core fattens with
decreasing values of $\beta$, therefore we define the length of a
string to be the number of lattice points belonging to the string
divided by the cross section of the corresponding ANO string. We will
discard blobs whose length is less than a cutoff 
(set to be order of the width of a string).

\section{Simulations} 

Previous numerical studies of SL strings have tested their
stability~\cite{Achucarro:1992hs} and formation rate in
two~\cite{Achucarro:1997cx} and three~\cite{Achucarro:1998ux}
dimensions, showing that formation rates can reach a third of the rate
of topologically stable strings at the lowest values of $\beta$
simulated (which was $\beta = 0.05$). Based on these studies, we work
under the assumption that the SL string network forms and we utilise
present high performance computing technology to achieve much longer
dynamical range to study the evolution of the network.  

The discretized field equations were derived from~(\ref{semilocalaction})
using techniques from Hamiltonian lattice gauge theories
\cite{Moriarty:1988fx} and solved using the staggered leapfrog
algorithm. Damping terms $\eta \dot\phi$ 
and $\eta \dot{A}_{\mu}$ are introduced
into the scalar and gauge field equations, respectively. 
We set the lattice spacing to be $\Delta$x $=1.0$ (the
string cores are at least three points wide) and the time step
$\Delta$t $=0.2$.

During the simulation total energy densities are calculated 
and strings are monitored via their magnetic energy. 
We study two main string observables as a function of time:
the distribution of strings as a function of length $n(l,t)$ and a
length scale $\xi(t) = \sqrt {V/L(t)}$ where $V$ is the volume and
$L(t) = \int_0^\infty n(l,t)l dl$ is the total string length. 
Linear growth of $\xi$ as a function of time is referred to scaling of 
the string network.
With periodic boundary conditions and our choice
of lattice spacing the time required for two
signals emitted from the same point and travelling in opposite directions
to interfere with each other is half of the side of the grid. 
This sets the maximum time the simulations can probe reliably.
If a string spanning the
box forms before that time, from the point of view of an observer in
the box it would appear coming from superhorizon scales.

We are interested in the regime $\beta \leq 1$. Fattening of the scalar 
core sets a lower bound $\sqrt{\beta}=0.2$ that we can examine numerically; 
for smaller values of $\beta$ scalar cores of 
different strings start to overlap in the initial configurations used. 
We tested several choices of $f$, the threshold of the magnetic field 
that decides if a point belongs to a string. 
For $0.2 \leq f \leq 0.4$ we do not observe any significant discrepancy 
in the length scale $\xi$ of the network.

Based on earlier works~\cite{Achucarro:1997cx,Urrestilla:2001dd} 
we do not expect the
formation of the network to be very sensitive upon the initial
conditions. Two different 
strategies for setting the initial conditions were considered:

a) One strategy consists of setting all fields (scalars and gauge) 
and gauge momenta to zero but giving the scalar field momenta some 
velocity obtained from a set of Gaussian random variables with zero 
mean. The configuration of uncorrelated velocities is then smoothed 
by averaging the velocities over nearest neighbours $s$ times. 

b) The second strategy is closer to the 
Vachaspati-Vilenkin simulations~\cite{Vachaspati:1984dz} where all 
field momenta an gauge fields are set to zero, but random phases 
are asigned to the scalar field, which in turn are smoothed out 
by averaging over neighbouring points $s$ times.

Note also that both types of initial configurations satisfy
Gauss' law. A set of simulations was performed in simulation boxes of 
size $384^3$ using both kinds of initial conditions varying the parameter $s$ 
from $10$ to $40$. 
We monitored the length scale $\xi$ and found that the simulations did not show
qualitatively different behaviour, and quantitatively, the deviations lay
within the statistical errors obtained by repeating the simulations
with several (of order ten) different initial configurations. Moreover, case
b) has typically enough energy so that the fields climb up the
potential and then the system goes to a state resembling that of case
a).

After some transient time, the system forms a
network of strings, and the network reaches a scaling regime. This
happens for all the initial conditions studied above, 
the only difference being the time it takes for the simulation
to go through the transient regime.
Therefore, we chose the most favourable way of creating initial
conditions from the point of view of reaching quickly a scaling regime. 
As the dynamical range of simulations is in any case fairly
limited, we seek to minimize the initial transient time. We do not
expect to capture all the physical phenomena in the phase transition
with this treatment, but neither is the main focus of this
study the precise description of the dynamics at the phase transition,
but rather the evolution and properties of the network, if formed and
persistent, after the phase transition.  For instance, the initial
conditions used are not thermal, but this is the case for e.g.
tachyonic preheating after inflation~\cite{Felder:2000hj}.

In the early universe there would be several sources of dissipation 
for the string networks, most notably the Hubble damping that 
decreases inversely proportional to time. In the case of our simulations there 
is the need to set the damping in such a way that the subsequent evolution of the 
network proceeds fast but without the fields overshooting. Several 
choices for $\eta$ were tested and for the data shown 
it is either a constant ($\eta = 0.05$),
or inversely proportional to time ($\eta=6/t$). During the first time
steps ($t<30$) the damping is set to $\eta=0.2$ to go through the
initial transient period fast and lose energy efficiently.  

\begin{figure}
\begin{center}
\includegraphics[width=0.94\hsize]{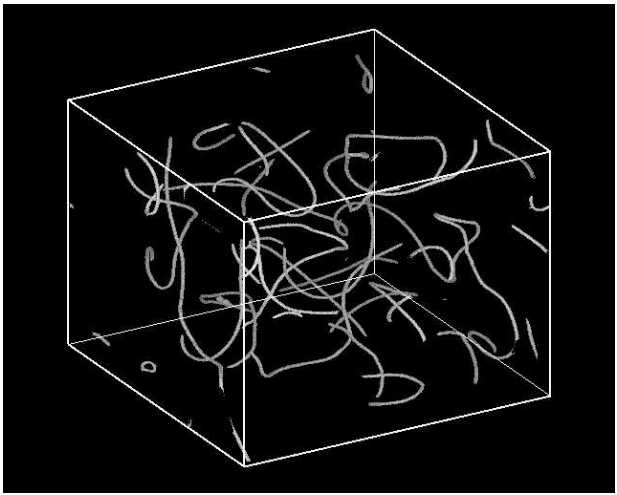}
\caption{\label{f:simulationbox} A snapshot of the simulation box
($512^3$) for $\sqrt\beta=0.2$, $\eta=0.05$ at $t=250$.}
\end{center}
\end{figure}

\section{Results}

Simulations in boxes size $512^3$ were performed for a set of ten
($\sqrt{\beta}>0.3$) or fifteen ($\sqrt{\beta}\leq0.3$) 
initial configurations for
each value of $\sqrt\beta$ ranging from 0.6 to 0.2 (we also verified the 
disappearance of the string network at the 
Bogomol'nyi limit, $\beta = 1$). We used initial
random velocities with smoothing 20 times ($s=20$), the
fraction $f=0.25$ for string identification 
and damping term either constant ($\eta=0.05$) or time dependent ($\eta=6/t$).

After the initial transient ($t\sim 30$) 
a network of string segments is formed. The
subsequent evolution of the segments depends strongly on the value of
$\beta$, but as a general rule strings are longer and their density increases 
as $\beta$ decreases. Intercommutations of strings are rare.

For $\sqrt{\beta}=0.6$ the maximum string length is around half 
the size of the simulation box. In principle, two strings of that length 
joining together could form a string spanning the box size, but we have 
not seen any such event, because the network is very sparse.
For $\sqrt{\beta}=0.4$ we have the first event of a string that spans the 
simulation box at time~$=250$. Decreasing $\beta$ even more 
leads to longer strings and a more dense network, 
for $\sqrt{\beta}=0.3$ approximately one third of the initial 
configurations lead to the formation 
of at least one percolating string; for $\sqrt{\beta}=0.2$ more than a half. 
Figure~\ref{f:simulationbox}
 shows a snapshot 
of the simulation box for $\sqrt{\beta}=0.2$, where there are several 
strings spanning the size of the simulation box. For $\sqrt{\beta}\leq0.25$ we find
some extremely long strings.  Analysis of the length distribution shows a qualitative change around this value of $\sqrt \beta \approx 0.25$.

\begin{figure}
\begin{center}
\includegraphics[width=0.94\hsize]{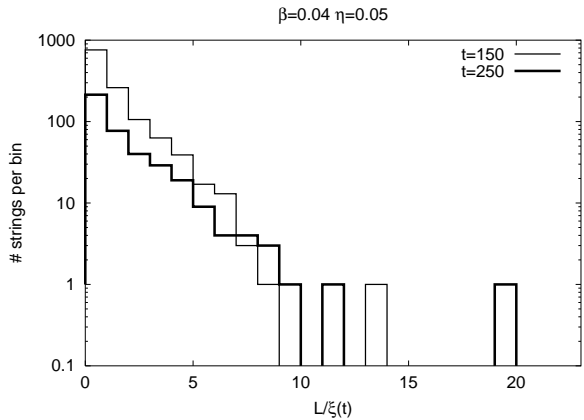}
  \caption{\label{f:distribution} The string length distribution
averaged over 15 simulations
for $\sqrt\beta=0.2$, $\eta=0.05$, at time = 150 (shaded) and 250
(thick line). Up to $ l/\xi \sim 10-13$ the segments lie on an
exponential distribution. The other points are extremely long strings
not belonging to the exponential. Those cases are only found for
$\sqrt{\beta}\leq0.25$.}
\end{center}
\end{figure}

For $\sqrt{\beta}>0.25$ the segments lie in an exponential distribution.
Nevertheless, for $\sqrt{\beta}\leq 0.25$ we find that besides the population of segments in the 
exponential distribution, we find extremely
 long strings ($l/\xi=19$ in Figure \ref{f:distribution}) which are clearly not part of the 
exponential. These  long strings percolate through the box and are the precursors of infinite strings at late times.

\begin{figure}
\begin{center}

\includegraphics[width=0.94\hsize]{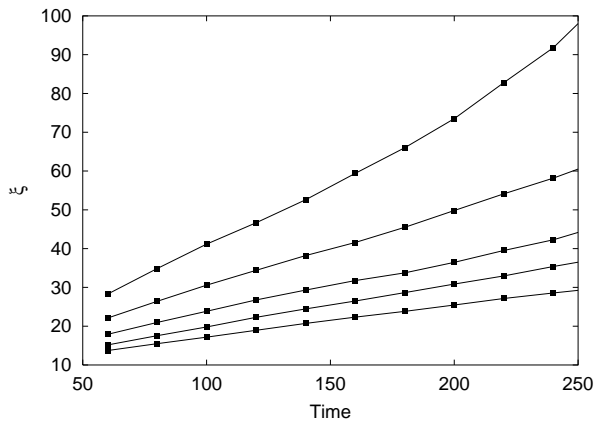}
  \caption{\label{f:scaling} $\xi(t)$ in a simulation box in a single run 
with $\sqrt{\beta}=0.6,0.5,0.4,0.3,0.2$ (from top to bottom), $\eta
\sim t^{-1}$.}

\end{center}
\end{figure}

The value of $\beta$ influences the network in two different ways. 
First, the initial
string density (after discarding the transient) grows with decreasing
$\beta$, see figure \ref{f:scaling}. Second, the accretion of disperse 
magnetic flux into tubes is much more efficient for
low $\beta$. To try to disentangle the two effects we have tested
configurations with different initial defect densities for low $\beta$
and found that they all quickly approach the same $\xi(t)$, suggesting
that flux accretion is the dominant effect that determines the scaling
behaviour.

When the string network becomes more dense, there are also
occasionally loops forming, but approximately only every third
simulation shows a loop for $\sqrt{\beta} \leq 0.4$.  The reason
might be simply an observational effect, loops disappear fast and are
not likely to be seen in snapshots. Moreover, we have not seen any
intercommutation in our simulations. On the other hand field theory
simulations do not tend to show many loops for ANO string networks
either \cite{Vincent:1997cx}. We do not observe a population of small
loops and such a tiny number of loops suggests that they cannot have a
major effect in the network.

Figure~\ref{f:scaling} shows the observable $\xi$ as a function of 
time in simulations where the damping term was set to be 
$\eta = 6/t$. The slow change in time for small $\beta$ proves 
that the string network 
can persist for a considerably long time and goodness of linear fit to data 
indicates a scaling network within the range we can probe.

Finally, we monitored the total energy densities in the simulation box. 
In all cases studied here the scalar gradients dominate the
energy density. This is not surprising in the
Bogomol'nyi limit, since the string network disappears, but it is the case 
also even at lowest values of $\beta$ we were able to probe. 
The approximate ratios of energy densities
(~kinetic~:~potential~:~magnetic~:~scalar gradients~) are $(2:1:1:15)$
for $\sqrt{\beta} =0.5$ and $(1:1:1:7)$ for $\sqrt \beta = 0.2$. 

\section{Discussion} 

Consider the exponential distribution of short segments.
At low $\beta$ the string segments are very light compared to the
(global) monopoles at the ends, which are strongly interacting and
playing substantial role in the evolution of the segments. 
This is just the opposite of the usual
situation encountered in cosmological scenarios in which the magnetic
monopole problem is solved by a subsequent phase transition in which
the magnetic flux of the (local) monopoles is confined to strings. In
this latter case the monopoles are very light and hardly contribute to
the dynamics, which are dominated by the string tension.

If a semilocal segment is much shorter than the average interstring
separation it is very likely that the segment will shrink
and the monopole and antimonopole at the ends will annihilate each
other. In this case one naively expects similar length distributions
whether the strings are heavy or light.
However, for long segments the difference becomes apparent. In
particular, the breaking of a semilocal string into a
monopole-antimonopole pair is strongly suppressed~\cite{Preskill:1992ck}. 
Tunneling is forbidden because there is no
configuration that the straight string can tunnel to, as the string
configuration has lower energy than the broken configuration. In this
respect, semilocal strings at low $\beta$ resemble topological
strings.  The break-up instability is the reason why in the usual
cases of (light) monopoles connected by strings, 
the length distribution stays exponentially supressed throughout. 
By contrast, in the semilocal case there is not such a 
suppression of long strings. Long SL
strings, once formed, are likely to continue growing by joining.

Thus it would be tempting to argue that, 
deep in the stability regime, a SL string
network could be considered effectively as a collection of global
monopoles joined by comparatively light strings. If this picture is
valid, the scaling in $\sigma$-model studies of global monopoles in an
expanding universe~\cite{Bennett:1990xy,Yamaguchi:2001xn} provides
further support for the scaling behaviour we observed within the
limited range we were able to probe. 
As time progresses, for sufficiently low $\beta$ we
expect the appearance of a population of infinite strings with typical
ANO behaviour, loop production, etc. but at a lower density than in
the topological case starting from the same initial state.

\section{Conclusions} 

In this paper we studied the evolution of 
semilocal string
networks deep in the stability regime using high performance
computing to carry out numerical
field theory simulations in flat space. Our main results are:

$\bullet$ A population of segments with an initially 
exponential length distribution. Short segments vanish rapidly and the 
mean string length grows. 

$\bullet$ Appearance of superhorizon strings (much larger than the box
size) percolating through the box deep in the stability regime,
$\sqrt{\beta} \lesssim 0.3$. For $\sqrt{\beta}\lesssim 0.25$ these
fast growing strings fall outside the exponential distribution.

$\bullet$ No significant population of small loops. In dense networks 
$\sqrt{\beta} \lesssim 0.4$
about one third of the 
simulations show a loop of a larger size.

$\bullet$ The string network can persist for a 
long time and for low $\beta$ the results are consistent with linear scaling.

$\bullet$ The energy density is dominated by scalar gradients.

These findings suggest that SL networks may have a cosmological
signature, in particular in the CMB and large scale structure, closer
to that of global defects (monopoles or textures) than to topological
strings. Current bounds~\cite{Durrer:2001cg} on global
textures in CMB allow about 13\% of the signal to come from defects at
$\ell = 10$.

The relevance of these
results to the cosmological scenarios
in~\cite{Urrestilla:2004eh,Dasgupta:2004dw} would require further study
in expanding backgrounds (see also \cite{Benson:1993at}) and the
development of analytic characterizations of the network to verify the
apperance of an infinite string population at late times, as was done
in ref. \cite{Copeland:1998na} for the case of dense networks of
cosmic string loops. 
Apart from the expansion, a complete
quantum field theoretical study of the conditions at 
formation is still lacking, as
pointed out in~\cite{Dasgupta:2004dw}.

If it is confirmed that the observational signal of the semilocal
network is similar to global textures or monopoles in the temperature
power spectrum \cite{BKHLU}, there exists a potential problem for
distinguishing between models via future data. A dedicated study is
needed to find out if there is any difference e.g. in polarisation of
vector modes, or in the possibility of gravitational waves radiating
from oscillating loops \cite{VS}.

\begin{acknowledgments}

The authors thank K. Benson, N. Bevis, M. Hindmarsh, A.R. Liddle,
C.J.A.P. Martins, K. Olum and A. Vilenkin for useful
conversations.  A.A. and P.S. are supported by the Netherlands
Organization for Scientific Research (N.W.O.) under the VICI
programme. P.S. was supported by Marie Curie Fellowship of the
European Community Program HUMAN POTENTIAL under contract
HPMT-CT-2000-00096.  J.U. is partially supported by the Spanish {\it
Secretar{\'{\i}}a de Estado de Educaci\'on y Universidades} and {\it
Fondo Social Europeo} and by the National Science Foundation. 
This work was partially supported by ESF COSLAB programme, FPA
2002-02037, 9/UPV00172.310-14497/2002 and FPA2005-04823.  We
 acknowledge the allocation of computing time on 
the UK National Cosmology Supercomputer funded by PPARC, HEFCE and
Silicon Graphics.

\end{acknowledgments}
%====================================================================

\end{document}